\documentclass[twocolumn,showpacs]{revtex4-1}
\usepackage{amsfonts}
\usepackage{amsmath}
\usepackage{graphicx}
\begin{document}
\draft

\title{Giant flux jumps through a thin superconducting Nb film }
\author{M.I. Tsindlekht$^1$, V.M. Genkin, I. Felner$^1$, F. Zeides$^1$, N. Katz$^1$, $\check{\text{S}}$.~Gazi$^2$, $\check{\text{S}}$. Chromik$^2$ }

\affiliation{$^1$The Racah Institute of Physics, The Hebrew University of Jerusalem, 91904 Jerusalem, Israel}

\affiliation{$^2$The Institute of Electrical Engineering SAS, D$\acute{u}$bravsk$\acute{a}$ cesta 9, 84104  Bratislava, Slovakia}

\begin{abstract}
The dynamics of magnetic field penetration into thin-walled superconducting niobium cylinders is experimentally investigated.
It is shown that magnetic field penetrates through the wall of a cylinder in a series of giant jumps with amplitude 10 - 20 Oe and duration of
a few $\mu$s. The jumps take place when the total current in the wall, not the current density, exceeds some critical value.
In addition there are small jumps and/or smooth penetration, and their contribution can reach $\simeq 20$ \% of the total penetrating flux.
It is demonstrated that the magnetic field inside the cylinder exhibits several oscillations. The number of giant jumps reduces with temperature.

\end{abstract}
\date{\today}
\maketitle

\section{Introduction}

The phenomenon of magnetic flux penetration into superconductors has longstanding scientific tradition.
The physics of this phenomenon in type II superconductors, for the magnetic field parallel to the sample surface, is simple.
For magnetic fields below $H_{c1}$ (in the Meissner state) the fields are totally expelled from the sample. Whereas above $H_{c1}$ the flux
penetrates via Abrikosov vortices. However in real superconductors, due to defects the vortex distribution is not uniform and in this case the
vortex lattice is in a metastable state. The metastable states of the vortices are the cause of numerous physical phenomena such as: flux creep,
magnetic relaxation~\cite{KIM,YOSI} and vortex avalanches, see review by Altshuler and Johansen and references therein~\cite{ALT}.
The explanation of the vortex avalanches was given mainly in the framework of two models: thermomagnetic instabilities \cite{MIN} and
self-organized criticality (SOC)  \cite{BAK}. The avalanche-like penetration of magnetic flux into a type II superconductor
was studied by numerous experimental methods. The most popular methods are magneto-optical imaging, micro-Hall-probe and
pick-up coil experiments~\cite{ALT}. Subsequently in this paper we will focus mainly on experiments with a pick-up coil.
The first pick-up coil experiments were carried out more than 50 years ago~\cite{BLANC,GOE}.
In these experiments voltage spikes that were induced in the pick-up coil by the magnetic moment jumps caused by a swept field
were measured.
The magnetic moment jumps at high field sweep rate, up to 1 kOe/s, were accompanied by increasing the sample temperature due to fast
motion of the vortex bundles \cite{BLANC,GOE}. Thermomagnetic instabilities are the cause of such heating \cite{MIN}.
However, at low sweep rates, the heating produced by a flux jump did not play a significant role \cite{HEID,FIELD}.
The experimentally observed avalanche size distribution, which indicates the number of vortices in a moving bundle, is an exponential \cite{HEID} or a
power-law \cite{FIELD} function of the avalanche size. The latter was used as a confirmation of SOC model. It was
demonstrated that the magnetic flux penetrates into the interior of the cylinder in two fashions. The first one is a smooth flux flow and the second one
is jump like. Contribution of a smooth flow was about 97\% of the total penetrated flux \cite{FIELD}.
Experiments reported in \cite{WIS,FIELD} were performed on hollow cylinders with large wall thickness $d\gg \lambda$,
where $d$ is a wall thickness and $\lambda$ is the London penetration depth.
In this case, as expected, avalanches were detected in magnetic fields higher than $H_{c1}$, ei.e. in a mixed state.
At the same time, we have shown that for a thin-walled cylinders of Nb, $d\leq \lambda$, an "avalanche"-like magnetic flux penetration
occurs not only in magnetic fields higher but in {\bf lower} than $H_{c1}$ of the film itself, i.e. in the vortex free regime \cite{Katz1}.
The latter phenomenon is very interesting, and the physical reasons for it are not yet clear.

In this paper, we present a comprehensive experimental study of the dynamics of magnetic flux penetration into thin-walled superconducting
Nb cylinders. The magnetic field penetrates through the walls of the cylinder in a series of giant jumps with magnetic field amplitude 10 - 20 Oe
and duration of a few microseconds. We have shown that magnetic field inside the cylinder exhibits several oscillations in each jump.
The jumps take place when the total current in the wall, not the current density, exceeds some critical value.
The giant jumps were observed in fields below and above $H_{c1}$ of the Nb film itself. In addition to the avalanche-like
penetration there is a smooth penetration and its contribution can reach about 20 \% of the total penetrated flux.

 \section{Experimental details}

\begin{figure}
\begin{center}
\leavevmode
\includegraphics[width=0.5\linewidth]{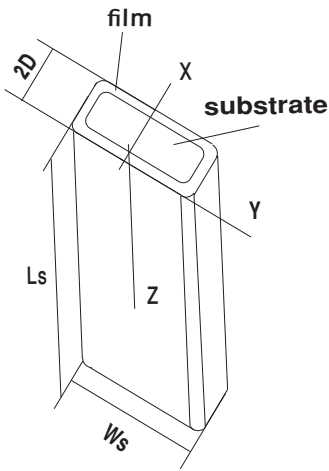}
\caption{Sketch of the rectangular sample. Here $\text{W}_s=3$ mm, and $2\text{D}=1.4$ mm are the substrate width and thickness,
respectively, with variable length, $\text{$L_s$}$.  }
\label{f1}
\end{center}
\end{figure}

Various thin-walled cylindrical samples with two cross-section shapes were prepared by dc magnetron sputtering of Nb on a rotated
sapphire and glass substrates at room temperature (Bratislava) and at $300\,^{\circ}$ C (Jerusalem). Sapphire substrates were manufactured
by Gavish Ltd company (Israel). The cross-section dimensions of the long parallelepiped sapphire substrates were 1.4 by 3 mm.
The corners of these substrates were rounded to a radius of 0.3 mm. A sketch of the rectangular sample is shown in Fig.~\ref{f1}.
The glass substrate has a circular cross-section with a diameter of 3 mm.  The lengths, film thicknesses, deposition temperatures and
labels of the samples are presented in the Table 1.

\vspace{5pt}
\begin{tabular}{|p{40pt}|l|p{60pt}|p{39pt}|p{50pt}|p{49pt}|}
\multicolumn{4}{c}{Table 1.}\\
  \hline
  Sample label & \textit{d} (nm) & Cross-section shape & Length (mm) & Deposition temp.

  ($\,^{\circ}$C)\\
  \hline
 F1 &60 &rectangular & 15& 300 \\
 \hline
  F8 &120 & rectangular & 15& 300 \\
  \hline
  F15 & 300 &rectangular & 15&300 \\
\hline
G7 & 100 & rectangular &7.5& 20 \\
\hline
G21 & 100 &rectangular& 21& 20 \\
\hline
G3a & 100 & circular & 7.5& 20 \\
 \hline

\end{tabular}
\vspace{5pt}

\begin{figure}
\begin{center}
\leavevmode
\includegraphics[width=0.8\linewidth]{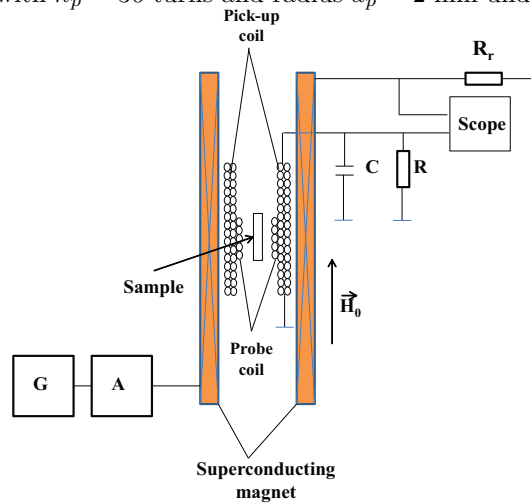}
\caption{(Color online) Block diagram of the experimental setup. Here {\bf C } is a capacitor of the coaxial cables, {\bf R} is a
shunt resistor for damping oscillations in a resonant circuit, {\bf G} is 33250A Agilent generator, {\bf A} is amplifier,
and {\bf R$_{\text{r}}$} is a current sensing resistor.}
\label{f2}
\end{center}
\end{figure}

The samples were inserted into a pick-up copper coil of length L=15 mm and N=2100 turns,
which was placed in a commercial superconducting quantum interference device (SQUID) magnetometer, MPMS5,
Quantum Design.
The magnetic field was directed parallel to the long axis of the sample. The superconducting magnet of the SQUID magnetometer was driven by an
arbitrary waveform generator, Agilent 33250A, and a "home build" amplifier. The applied field was slowly changed with a rate of: 4, 8, or 16 Oe/s
from zero to 1.86 kOe and then back to zero. The voltage spikes induced by a swept magnetic field were measured
by InfiniiVision, DSO-X 3024A, Agilent oscilloscope. The oscilloscope second channel was used for magnetic field measurement.
Fig. \ref{f2} shows a block diagram of the experimental setup.

\section{Pick-up coil calibration}

\begin{figure}
\begin{center}
\leavevmode
\includegraphics[width=0.8\linewidth]{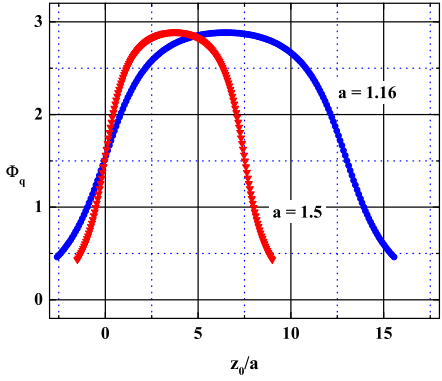}
\caption{(Color online) $\Phi_q$ versus $z_0$.}
\label{f3}
\end{center}
\end{figure}

To calibrate the pick-up coil we wound a small probe coil with $n_p=30$ turns and radius $a_p=2$ mm under the pick-up coil and
drove it by a current jump. A magnetic flux $\Phi_1$ induced in one turn of the pick-up coil (radius $R$) by a
linear circular current $J$ with radius $a$ is~\cite{LL}
 \begin{equation}\label{Eq1}
 \Phi_1=\frac{4\pi J}{c}\int_0^\pi\frac{a\cos(\varphi)d\varphi}{(a^2+R^2+z^2-2aR\cos(\varphi))^{1/2}},
\end{equation}
where $z$ is the distance between these two circles which are parallel to each other. From this expression we get magnetic flux
through the pick-up coil
\begin{equation}\label{Eq2}
 \phi=\frac{4\pi a^2 JN}{cL}\Phi_q(R/a,z_0/a,L/a),
\end{equation}
where
\begin{widetext}
\begin{equation}\label{Eq3}
 \Phi_q(r,z,l)=r\int_0^\pi d\varphi\cos(\varphi)\ln \Big(\frac{(1+r^2 +(l-z)^2-2r\cos(\varphi))^{1/2}+(l-z)}
 {(1+r^2 +z^2-2r\cos(\varphi))^{1/2}-z}\Big)
\end{equation}
\end{widetext}
is dimensionless function and $z_0$ is $z$-coordinate of the circular current. The $Z$-axis is chosen along the pick-up coil.
The coordinates of the pick-up coil's ends are $0$ and $L$.
 Fig.~\ref{f3} shows $\Phi_q(z_0)$ for two values of radius $a$, 1.16 and 1.5.
For $n_p$ turns of the probe coil of radius $a_p$ and length $l_p\ll L$, located in a center of the pick-up coil, $z_0 = L/2$, we obtain
\begin{equation}\label{Eq4}
 \phi=n_p\frac{4\pi a_p^2 JN}{cL}\Phi_q(R/a_p,L/a_p,L/a_p).
\end{equation}

Now we assume that a thin-walled circular hollow tube with length $L_s$, radius $a_t$, and wall thickness $d\ll a_t$ is inserted symmetrically
into the pick-up coil
and $\Delta H$ is the difference between magnetic fields inside and outside the tube. In this case the surface current density in the
wall is $j=c\Delta H/4\pi$. Flux that this current creates in the pick-up coil is
\begin{widetext}
\begin{equation}\label{Eq5}
 \phi=\frac{c\Delta H}{4\pi}\frac{4\pi a_t^2 JN}{cL}\int_{\Delta z}^{\Delta z+L_s}dz\Phi_q(R/a_t,z/a_t,L/a_t)=
 \frac{c\Delta H}{4\pi}\frac{4\pi a_t^3N}{cL}\Phi_t,
\end{equation}
\end{widetext}
where $\Delta z$ is a distance between the edges of the tube and pick-up coil. The $\Phi_t$ can be calculated by a numerical integration of
the data presented in Fig.~\ref{f3}. Equation~(\ref{Eq5}) for a circular tube is applied for a rectangular tube with the same cross section area .

The voltage jump in a pick-up coil due to the change of magnetic flux $\phi$ is
\begin{equation}\label{Eq6}
 V(t)=\int_0^td\tau K(t-\tau)\frac{d\phi(\tau)}{d\tau},
\end{equation}
where $K(t)$ is the response to the $\delta$-like flux jump. The integral
\begin{equation}\label{Eq7}
\overline{V}=\int_0^{\infty}V(t)dt=\Delta\phi\int_0^{\infty}K(t)dt
\end{equation}
is proportional to the total change of magnetic flux in the pick-up coil.

\begin{figure}
\begin{center}
\leavevmode
\includegraphics[width=0.8\linewidth]{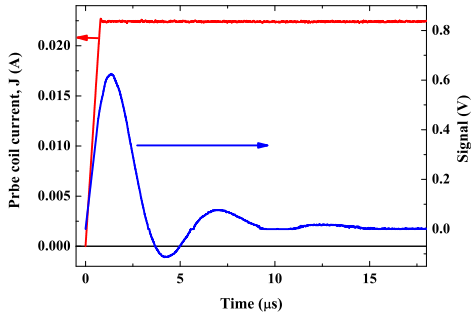}
\caption{(Color online) Current in the probe coil and the voltage on the pick-up coil as a function of time. }
\label{f4}
\end{center}
\end{figure}

The change of magnetic field inside a tube can be found from the time dependant voltage
in the pick-up coil and the response to the current jump in the probe coil from
\begin{widetext}
\begin{equation}\label{Eq8}
\Delta H=120\pi J\frac{\overline{V}_s}{\overline{V}_p}
\frac{a_p^2}{a_t^3}\frac{\Phi_q(R/a,L/2a,L/a)}{c\Phi_t}=\\
D\frac{\overline{V}_s}{\overline{V}_p}.
\end{equation}
\end{widetext}
Parameter \textit{D}
in Eq. (\ref{Eq8} was calculated and it is approximately equal 2.2 for samples F1, F8, F15 and G3a, 1.8 and 3.7 for samples G21 and G7, respectively. We took into account that the pick-up coil has several layers.
Fig.~\ref{f4} shows time dependences of the current in the probe coil and the voltage signal in the pick-up coil.

The data presented in Fig. \ref{f4} allow us to trace the time dependence of the magnetic flux, $\phi(t)$. Laplace transform of Eq.(\ref{Eq6}) yields
\begin{equation}\label{Eq9}
V(s)=K(s)\dot{\phi}(s).
\end{equation}
Using this equation for data that were obtained with the probe coil, Fig. \ref{f4}, during a jump, the time derivative of the magnetic flux is
\begin{equation}\label{Eq10}
\dot{\phi}(t)=\frac{1}{2\pi i}\int ds \exp(st)V(s)\dot{\phi}_p(s)/V_p(s),
\end{equation}
where $V(s)$ and $V_p(s)$ are the Laplace transform of the signal $V(t)$ during the jump and the signal $V_p(t)$ in the pick-up coil due to the
current jump in the probe coil. $\dot{\phi}_p(s)$ is the Laplace transform of the time derivative of the magnetic flux that the probe coil creates,
$\dot{\phi}_p(s)\propto dJ(t)/dt$, where $J(t)$ is a current in a probe coil. All integrals can be calculated with fast Fourier transform.

\section{Experimental results}
\begin{figure}
\begin{center}
\leavevmode.
\includegraphics[width=0.8\linewidth ]{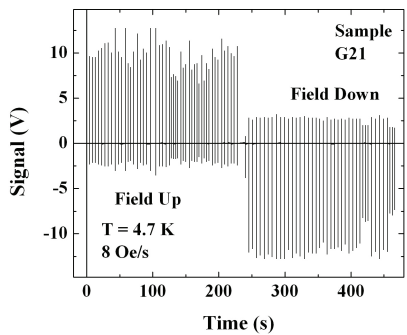}
\caption{Sequence of pulses in sample G21 at T = 4.7 K and seep rate 8 Oe/s.}
\label{f5}
\end{center}
\end{figure}

\begin{figure}
\begin{center}
\leavevmode
\includegraphics[width=0.9\linewidth]{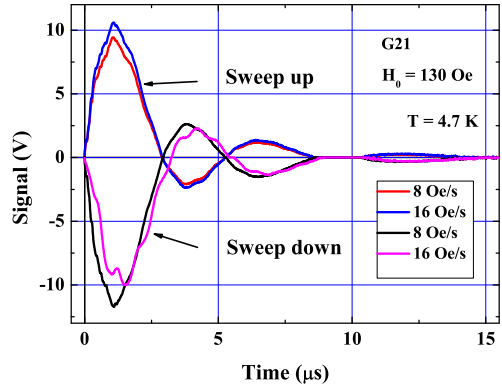}
\includegraphics[width=0.9\linewidth]{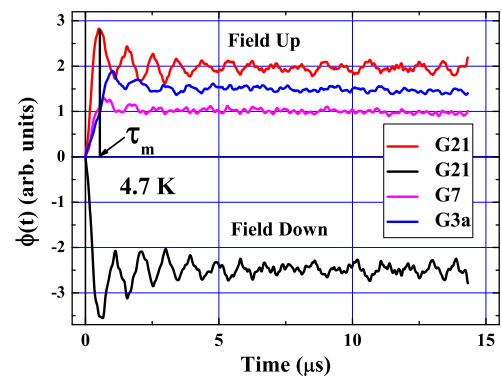}
\caption{Upper panel: Signals at $H_0=130$ Oe in ascending and descending fields that was observed in sample G21.
Lower panel: Magnetic flux $\phi(t)$ as a function of time for the 4$^{th}$ jump at 8 Oe/s ($H_0\approx 130$ Oe)
that was observed in sample G21 and the 1$^{th}$ jumps for G7 and G3a samples ($H_0\approx 30$ Oe ) at 4 Oe/s.}
\label{f6}
\end{center}
\end{figure}

\begin{figure}
\begin{center}
\leavevmode
\includegraphics[width=0.9\linewidth]{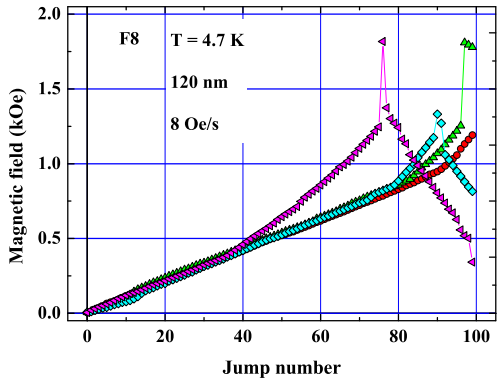}
\caption{(Color online) Magnetic field at which the jump was observed, $H_n$, as a function of jump number.
Subsequent measurements were repeated and done after ZFC.}
\label{f7}
\end{center}
\end{figure}

Fig. \ref{f5} presents the voltage pulse sequence for sample G21 at 4.7 K. The sample was zero field cooled (ZFC). Then the field was slowly
ramped with a rate of 8 Oe/s to 1.86 kOe and then swept back to zero.
Each vertical line in Fig. \ref{f5} is actually the time dependent voltage signal as shown in upper panel of Fig. \ref{f6}. The signal changes its
polarity when the magnetic field starts to decrease. Increasing the rate to 16 Oe/s did not cause any considerable change. From the obtained data
using Eq. (\ref{Eq10}) the variation  of magnetic flux, $\phi$ which penetrates into the cylinder, with time is deduced as shown in lower panel of
Fig. \ref{f6}. Unexpectedly, flux oscillations during the jump are observed. It is clear from Eq. (\ref{Eq10}) that these oscillations are not manifestation of the electric circuit response
(upper panel of Fig. \ref{f6}), as discussed above. Similar results were obtained for all other samples. Note that slowly ramping magnetic
field induces a small voltage on a pick-up coil due to smooth flux or small jumps' penetration.
This small voltage cannot be detected by our setup. The magnetic field $H_n$ at which the n$^{th}$ jump was detected is a random quantity.
Fig. \ref{f7} demonstrates few subsequent measurements of $H_n$ at 4.7 K for F8 sample after ZFC at each run.
Value of $H_n$ and the total number of jumps changes from one run to another.

\begin{figure*}

\begin{center}
\hbox{
\includegraphics[width=0.33\linewidth]{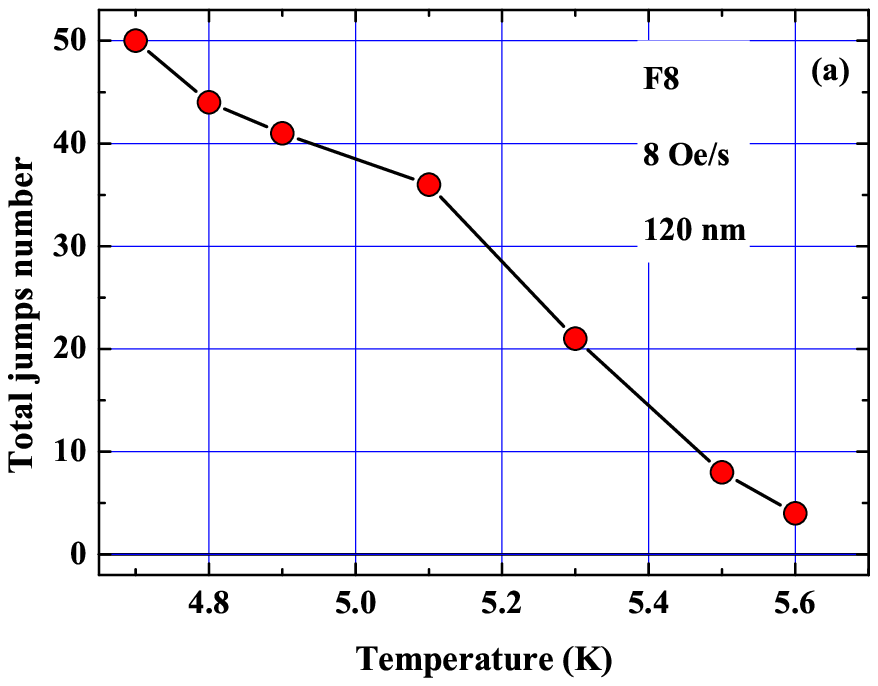}

\includegraphics[width=0.33\linewidth]{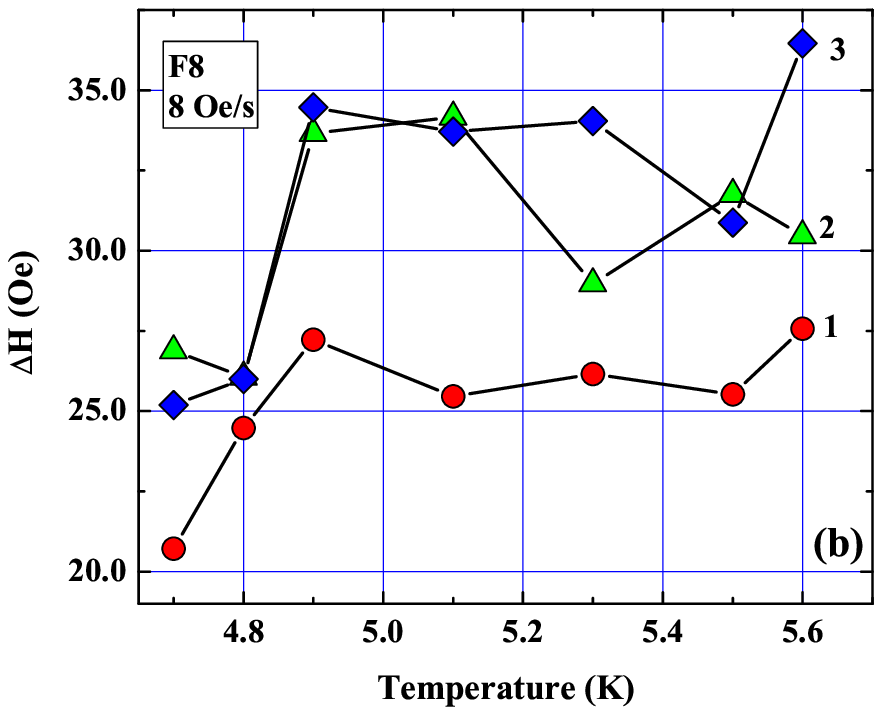}
\includegraphics[width=0.33\linewidth]{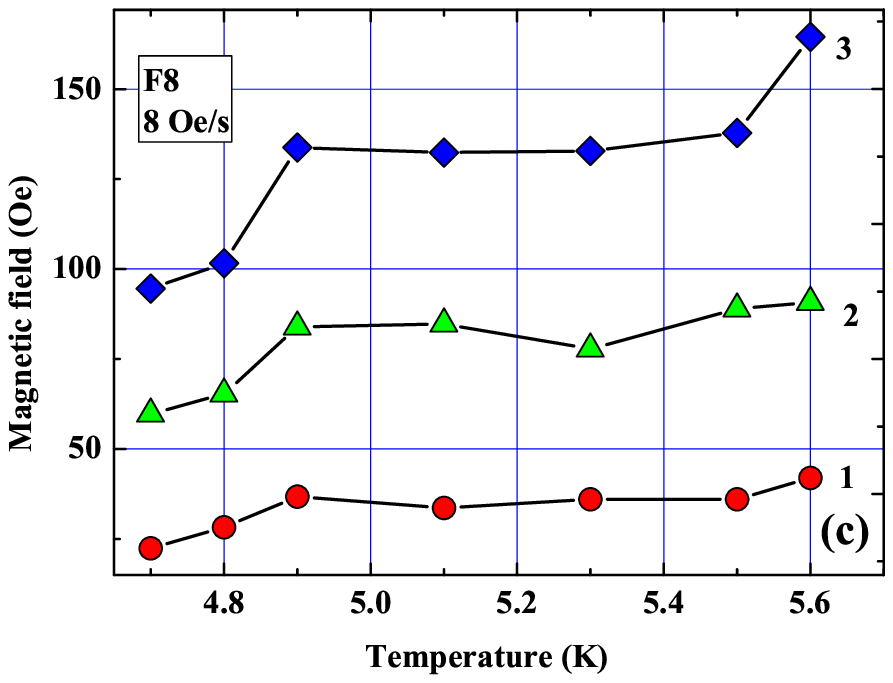}
}

\caption{(Color online) Panel \textit{a}: The total jump number in ascending field versus temperature.
Panel \textit{b}: Temperature dependence of the first three field jumps, $\Delta H_n$, see Eq.(\ref{Eq8}). Panel \textit{c}:
Magnetic field of the first three jumps as a function of temperature. The number near the curves in panels {\textit b} and {\textit c}
is the jump number.
The measurements presented in all three panels were obtained by ascending the magnetic field with a sweep rate of 8 Oe/s.}

\label{f8}
\end{center}

\end{figure*}

\begin{figure}
\begin{center}
\leavevmode
\includegraphics[width=0.8\linewidth]{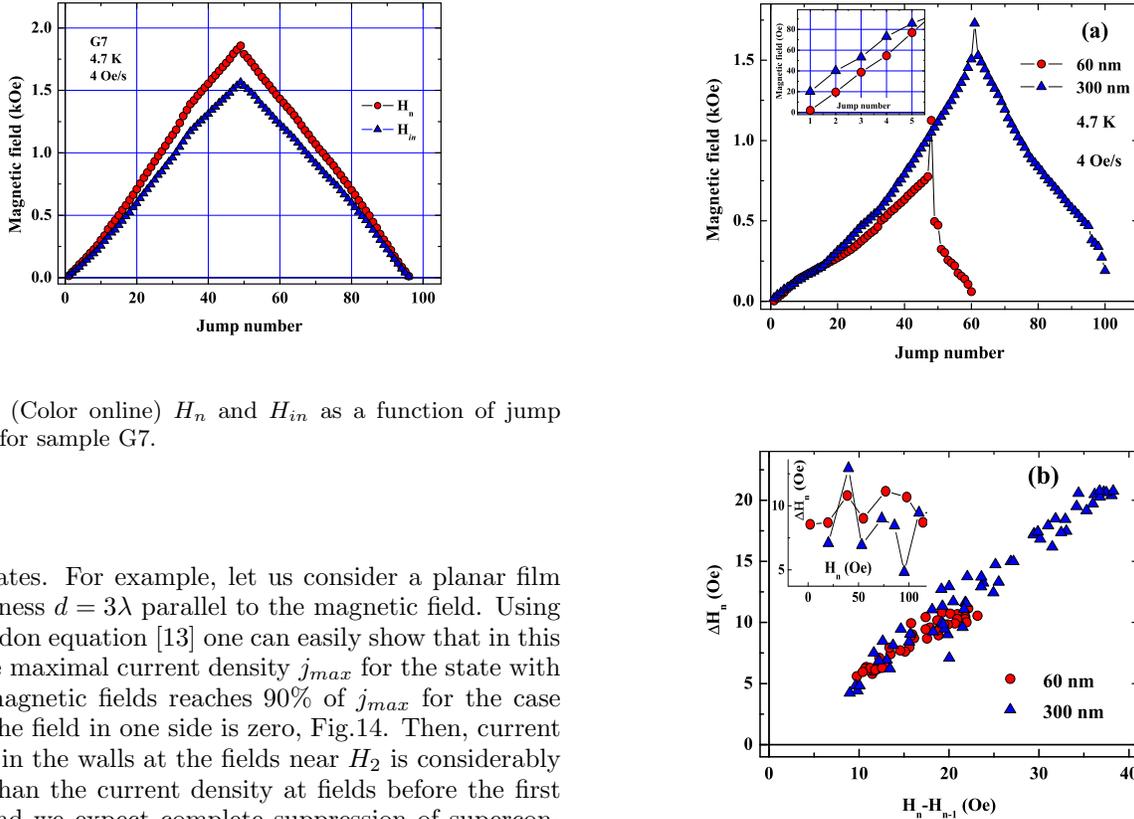}
\caption{(Color online) $H_n$ and $H_{in}$ as a function of jump number for sample G7.}
\label{f8d}
\end{center}
\end{figure}

The total number of jumps dramatically decreases with temperature and they completely disappear near 5.6 K, as shown in Fig. \ref{f8} panel $a$.
The temperature dependences of $\Delta H_n$ (Eq. (\ref{Eq8})) and $H_n$ for the first three jumps of sample F8 are presented in Fig. \ref{f8}
panels  $b$ and $c$, respectively.
The first jump at 4.7 K occurs in field $H_1 \approx 21$ Oe (Fig. \ref{f8}(c)) and at this jump $\Delta H_1 \approx 21$ Oe, Fig. \ref{f8}(b).
Thus, after the first jump the field inside the cylinder equals the external one. Estimation of the relative role of large jumps on the
flux penetration can be obtained from data shown in Fig. \ref{f8d}. This figure shows the calculated field inside a cylinder after {\textit n}
jumps, $H_{in}=\sum_{i=1}^{n}\Delta H_i$,
for sample G7 as a function of the jump number assuming that all penetrations are via large jumps only. Here $\Delta H_i$ is the magnetic field
jump at the i$^{th}$ event. The maximal value after 49 jumps of $H_{in}$ for sample G7 is $\approx 1.57$ kOe - average value of $\Delta H$
per one jump is about 32 Oe. The average magnetization of sample G7 at $H_0\approx1.86$ kOe is $m \approx -2$ emu/cm$^3$.
The magnetic induction inside the cylinder $B=H_0+4\pi m \approx 1.83$ kG, and the average $\Delta H$ per one jump
is $\approx 37$ Oe. It means that only $\approx 5$ Oe per jump penetrate through the wall via small jumps and/or smooth flux penetration.
These data show that the field penetrates through the walls via large jumps and the contribution of other undetected small jumps and/or smooth
penetration are small. This is in contrast with reported contribution 97\% for fields above $H_{c1}$ in a thick-walled cylinder~\cite{FIELD}.

\begin{figure}
\begin{center}
\leavevmode
\includegraphics[width=0.8\linewidth]{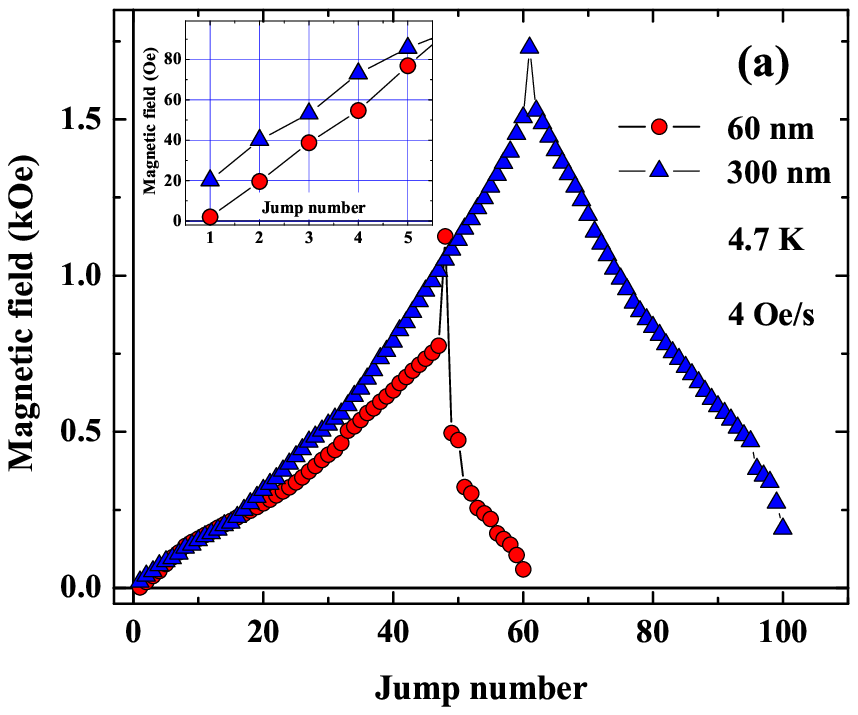}
\includegraphics[width=0.8\linewidth]{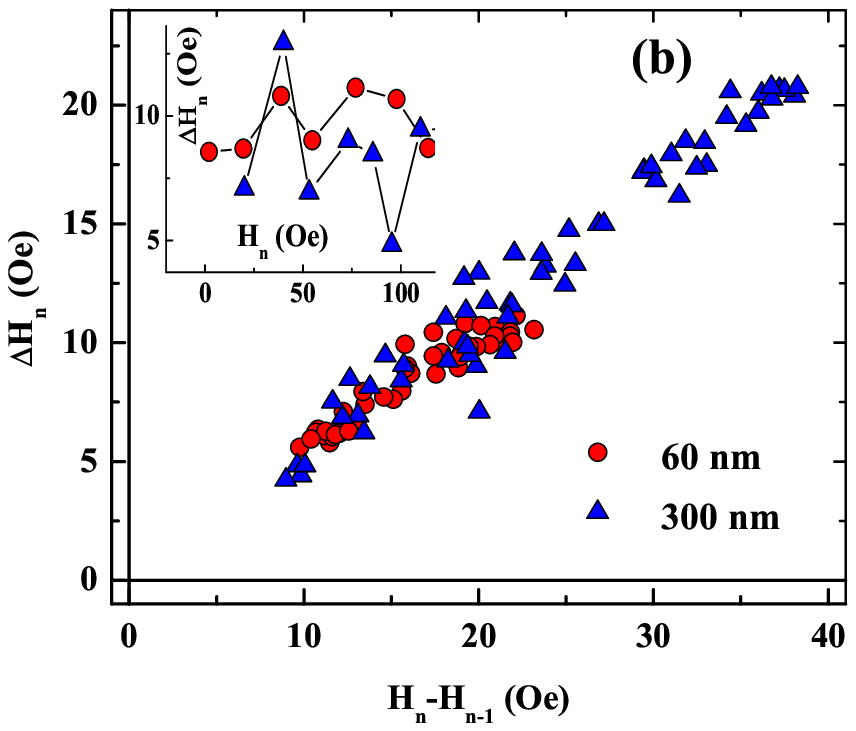}
\caption{(Color online) Panel \textit{a}: Magnetic field of jump as a function of jump number for F1 and F15 samples.
The first 100 jumps are shown only. Expanded view for the first five jumps is presented in inset.
Panel \textit{b}: $\Delta H_n$ as a function of $H_n-H_{n-1}$ for F1 and F15 samples.
Inset shows $\Delta H_n$ as function of a magnetic field. }
\label{f9}
\end{center}
\end{figure}

Fig. \ref{f9}(a) depicts $H_n$ as function of the jump number $n$ for samples F1 and F15 with wall thickness $60$ and $300$ nm, respectively.
The jumps for sample F1 disappear when the applied field exceeds 1.1 kOe, whereas for sample F15 the last jump was observed around
1.5 kOe. The average difference between $H_n$ of neighboring $dH_{av}=\langle H_n-H_{n-1}\rangle$ in ascending field equals
$\approx 17$ and $\approx 25$ Oe  for samples F1 and F15, respectively. For the first 20 jumps at low fields,
$H_0\leq 320$ Oe, the $dH_{av}$ decreases to $\approx 14$ and 16 Oe for samples F1 and F15, respectively.
The inset to Fig.\ref{f9}(a) shows an expanding view of $H_n$ versus $n$ at low fields.
The magnetic field that penetrates during each jump $\Delta H_n$ depends 
on the change in field from the previous jump, $H_n-H_{n-1}$,
as it is shown Fig. \ref{f9}(b) for samples F1 and F15.
Inset to Fig. \ref{f9}(b) presents $\Delta H_n$ as a function of $H_n$  for several first jumps.
These data are scattered and the difference between these two sets is not essential regardless the fivefold difference in the walls thickness.

\begin{figure}
\begin{center}
\leavevmode
\includegraphics[width=0.8\linewidth]{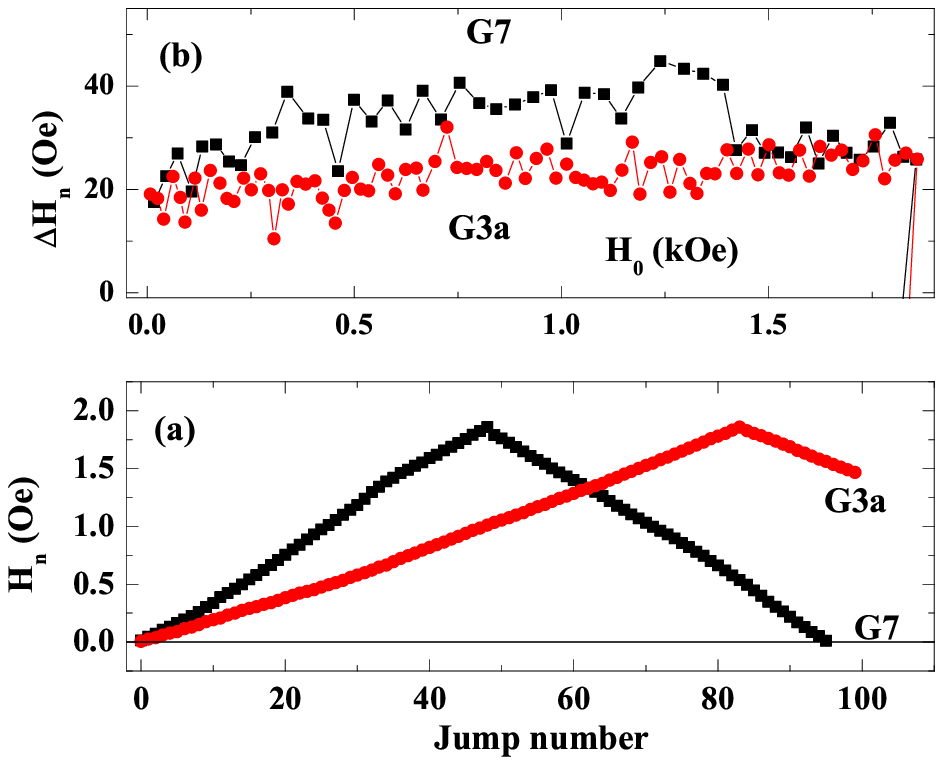}
\caption{Panel \textit{a}: $H_n$ as a function of jump number for samples G3a and G7 at 4.7 K. Panel \textit{b}:
Field dependence of $\Delta H_n$ for samples G3a and G7.}
\label{f10}
\end{center}
\end{figure}

\begin{figure}
\begin{center}
\leavevmode
\includegraphics[width=0.8\linewidth]{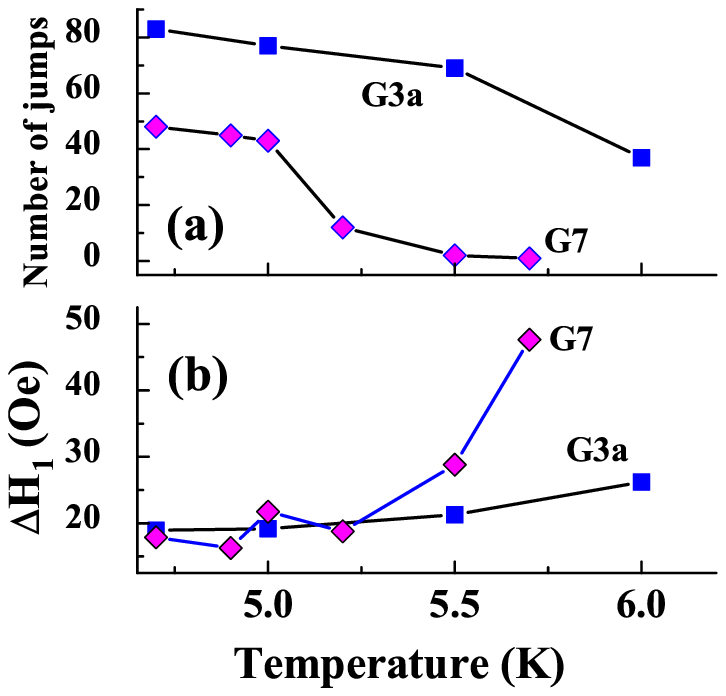}
\caption{Panel \textit{a}: The total number of jumps for samples G3a and G7 versus temperature in ascending magnetic field.
Panel \textit{b}: $\Delta H_1$ as a function of temperature for G3a and G7 samples.}
\label{f11}
\end{center}
\end{figure}

Until now we have presented the data for samples with rectangular cross section.
To rule out of the possibility that the observed features are due to the sample corners, where the deposited film may
have different properties, we show the results obtained for a sample with a circular cross section.
Fig. \ref{f10}(a) shows $H_n$ as a function of $n$ for samples G3a and G7 at 4.7 K.
The average field distance between adjacent jumps, $dH_{av}$, is lower for sample G3a and then amplitude of the field jump
$\Delta H_n$ is also lower than for sample G7, Fig. \ref{f10}b.
The total number of jumps in ascending field as a function of temperature for  G7 and G3a samples is presented in Fig. \ref{f11}\textit{a}.
Temperature dependences of $\Delta H_1 (T)$ for samples G3a and G7 are demonstrated in Fig. \ref{f11}(\textit{b}).
The comparison of two the sets of experimental data shows that there are quantitative differences between samples G3a and G7,
the behavior is similar and the role of the film deposited on rounded corners is not essential.

\section{Discussion}

\begin{figure}
\begin{center}
\leavevmode
\includegraphics[width=0.8\linewidth]{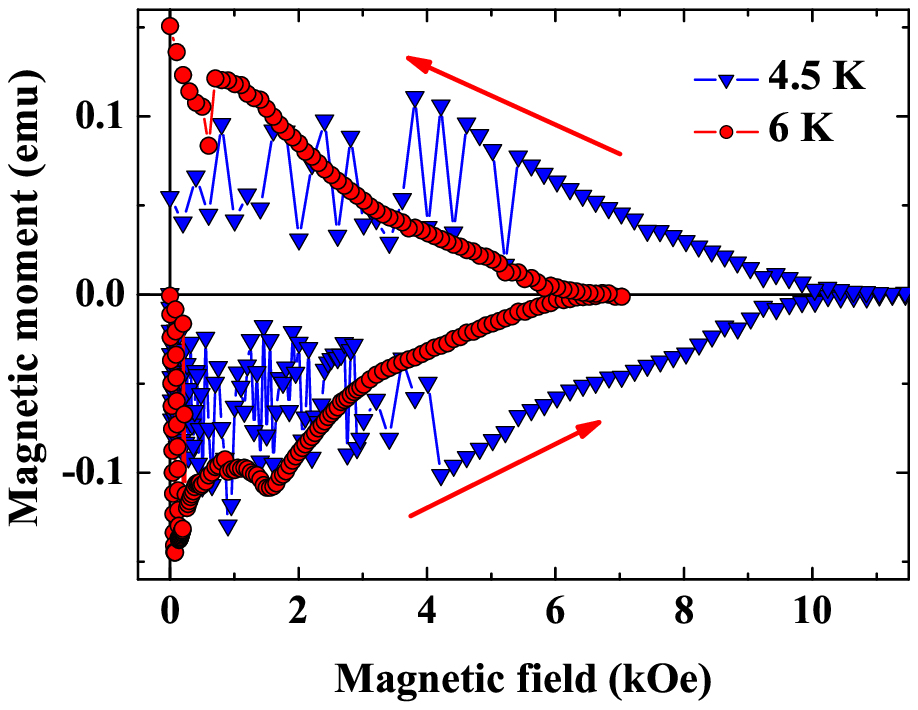}
\includegraphics[width=0.8\linewidth]{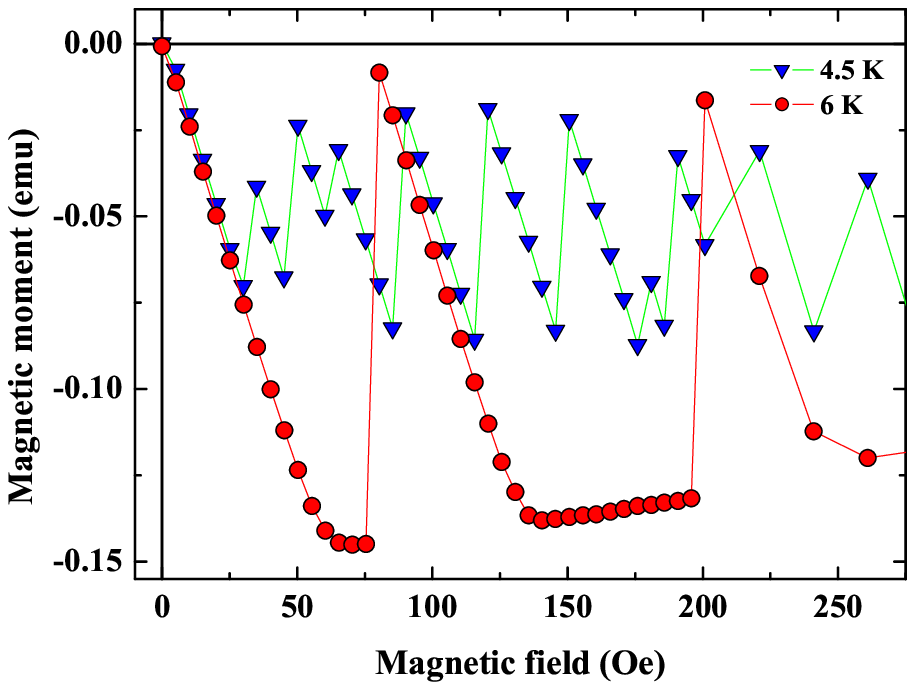}
\caption{Upper panel: Magnetization curves of G7 sample at 4.5 and 6 K after ZFC .
Lower panel: Expanded view of magnetization curves at low magnetic fields. }
\label{f12}
\end{center}
\end{figure}

The most striking phenomenon observed here is the magnetic field "avalanche"-like penetration into thin-walled
cylinder under external fields below $H_{c1}$. Direct determination of $H_{c1}$ for thin-walled cylindrical samples is impossible
due to magnetic moment jumps at low fields. However, the estimation of $H_{c1}$ can be done using magnetization curves
of a planar film. $H_{c1}$  was defined by the deviation from linearity in the magnetization curve at low fields for a planar film with
thickness 240 nm. It was found that $H_{c1}\approx 350$ Oe at 4.5 K and "avalanche"-like jumps of magnetic moment at low fields
are absent. Although this method is not accurate, we observed jumps of the magnetic moment
into cylindrical samples at fields of 20-40 Oe, which are significantly lower than the $ H _{c1} $.
Fig. \ref{f12} shows magnetization curves of the G7 sample at 4.5 and 6 K (upper panel). Lower panel of this figure demonstrates expanded view
of these curves at low magnetic fields. One can see that after the jump the superconducting state is restored, Fig. \ref{f12}, lower panel.
Above a specific jump at low fields the magnetic moment is a linear function of a
magnetic field with $dM/dH_0$ that corresponds to complete screening.

\begin{figure}
\begin{center}
\leavevmode
\includegraphics[width=0.8\linewidth]{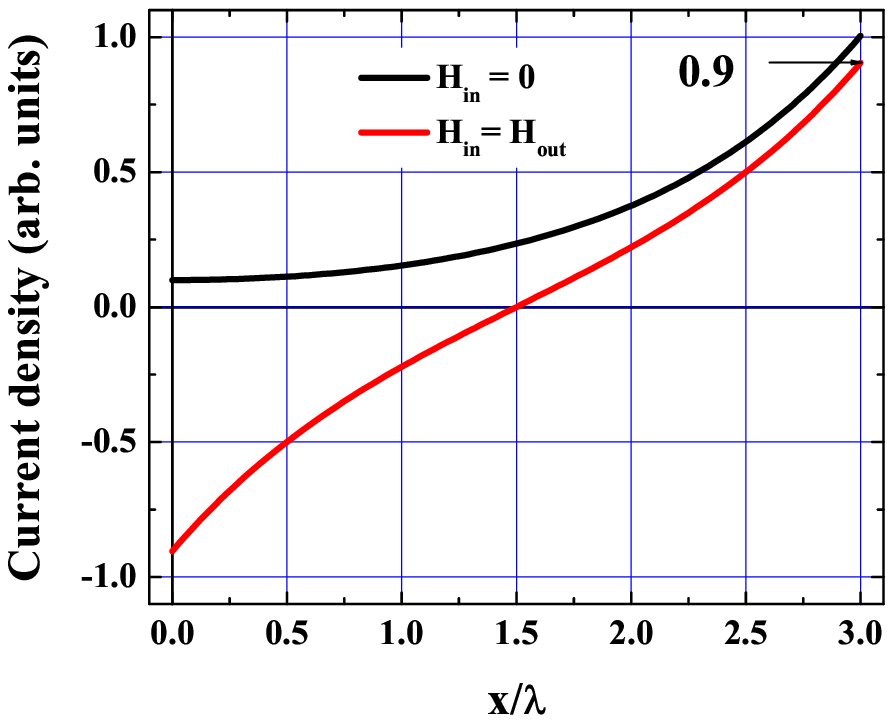}
\caption{(Color online) Current density in a film with thickness $3\lambda$ as function of coordinate for two cases,
first $H_{in}=H_{out}$ - red line and second $H_{in}=0$ - black line.}
\label{f13}
\end{center}
\end{figure}

Possible superconducting states of a hollow cylinder are: (i) metastable state with the internal magnetic field different from the external one and
non-zero total current in the wall, and (ii) stable state with equal magnetic fields and zero total current in the wall.
However, the maximal current density in the wall is actually the same for both states.
For example, let us consider a planar film of thickness $d=3\lambda$ parallel to the magnetic field. Using the London equation \cite{PG}
one can easily show that in this case the maximal current density $j_{max}$  for the state with equal magnetic fields reaches 90\% of $j_{max}$
for the case where the field in one side is zero, Fig.\ref{f13}.
Then, current density in the walls at the fields near $H_2$  is considerably larger than the current density at fields
before the first jump and we expect complete suppression of superconductivity in fields far below $H_2$ .
However,  Fig. \ref{f8} and Fig. \ref{f12} show that current density has no essential role and jump occurs when the total current in the
walls exceeds some critical value. If the field penetrates by these large jumps then the total current in
the walls before the n$^{th}$ jump has to be proportional to $H_n-H_{n-1}$ as it was observed for both samples F1 and F15.
Fig. \ref{f9}(\textit{b}) shows that the $\Delta H_n$ is indeed approximately proportional to $H_n-H_{n-1}$, despite
the fivefold difference in the sample wall thickness.

\begin{figure}
\begin{center}
\leavevmode
\includegraphics[width=0.8\linewidth]{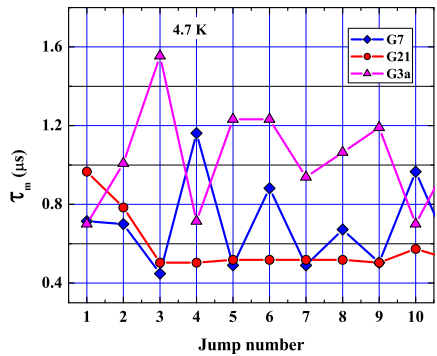}
\caption{(Color online) $\tau_m$ as a function of jump number for G7, G21 and G3a samples.}
\label{f14}
\end{center}
\end{figure}

We consider the possibility that the physical reason of these jumps is an instability at the end faces of the cylinder.
The magnetic fields at the sample end have components normal to the wall, $H_r$, as well as parallel to it. For a circular
hollow cylinder of radius $r$ and
length $L$ with constant azimuthal current density $J_{\varphi}(z)=J_0$ in the wall $H_r(z/r)=F((z-L)/r)-F(z/r)$ where
\begin{equation}
F(z)=\frac{J_0}{c}\int_0^{\pi}\frac{\cos(\varphi)d\varphi}{(1+z^2/2-\cos(\varphi))^{1/2}},
\nonumber
\end{equation}
as follows from Eq. (\ref{Eq1}) for $R=a$. We have $F(z) \propto \ln(z)$ when $z\rightarrow 0$. The Lorenz force
$F_l\propto J_{\varphi}H_r$
pushes the magnetic force lines into the cylinder on one end and pulls them out on the other end.
As result, the magnetic field in the cylinder increases while the current in the wall decreases.
In this case the rise time for magnetic field inside the cylinder, $\tau_m$, (see Fig. \ref{f6}b) may be dependent on sample length.
Fig. \ref{f14} shows  $\tau_m$ for the samples G3, G7 and G21 as function of the jump number.
In spite of threefold difference between the sample lengths  $\tau_m$ is almost the same.
We should mention that the role of the end faces in thin-walled superconducting cylinders under high frequencies was already discussed many
years ago \cite{TUL}.

The jumps disappear with temperature, Fig. \ref{f8}(\textit{a}) and Fig. \ref{f12}. This resembles thermomagnetic instability frequently
observed in superconductors \cite{MIN,ALT}. The thermomagnetic scenario requires presence of vortices in the walls.
In an infinite film, in parallel magnetic field, vortices should only exist at fields larger than $H_{c1}$ which is much higher than fields at
which jumps were observed.
We suggest that in low fields vortices penetrate into a wall due to the normal component of magnetic field at the end faces.
Indirect conformation of this is the similarity between jumps at small and large fields when vortices definitely exist in the walls.
However, the field of the first jump $H_1$ is increasing with temperature whereas below $H_1$ $dM/dH_0$ does not depend on temperature.
This is shown for sample G7 at 4.5 and 6 K in Fig. \ref{f12}. Thermomagnetic instability implies that the critical current decreases with
increasing temperature. Here we face the opposite case. At 6 K the instability or first jump takes place for a larger total current in the walls
than at 4.5 K. We would like to note that the standard approach to thermomagnetic instability \cite{MIN,ALT} is not applicable to the circumstances
of our experiment. The theory requires the existence of the Bean critical state in the film \cite{BEAN}.
But in a thin film with thickness equal to the size of one or two vortices the Bean critical state cannot arise.

\section{conclusion}

We have studied the dynamics of flux penetration into thin-walled superconducting niobium cylinders.
It was shown that magnetic flux penetrates through the walls in a series of giant jumps with duration of a few microseconds.
In addition, there is another flux penetration mechanism which contributes $\approx 20$ \% to the total
penetrated flux. The magnetic field inside the cylinder exhibits several oscillations during each jump.
The current density in the wall does not play an essential role in the observed phenomena. Jumps start when the total current
in the wall exceeds some critical value. The jumps were observed at a temperature of 4.5 K and completely disappeared at $\approx  7$ K.
Such behavior resembles thermomagnetic instability of vortices but it was observed in fields below $H_{c1}$ of the films, i.e. in
a vortex-free state. We consider the possibility that the instability at the end faces of the sample could be the source of these jumps because the
magnetic field at the sample end faces has a component normal to the wall.
The influence of the end faces on the flux jumps in such samples has to be studied using a local probe technique.

\section{acknowledgments}
We thank J. Kolacek, P. Lipavsky and V.A. Tulin for fruitful discussions.
This work was done within the framework of the NanoSC-COST Action MP1201.
Financial support of the grant agency VEGA in projects nos. 2/0173/13 and 2/0120/14 are kindly appreciated.

\end{document}